\documentclass[aps,epsf,preprint,showpacs,superscriptaddress]{revtex4}
\usepackage{amsmath}
\usepackage{epsfig}

\begin{document}

%
\title{Thermopower of a superconducting single-electron transistor}

\author{Marko Turek}
\email{marko.turek@physik.uni-regensburg.de}
\affiliation{Institut f{\"u}r Theoretische Physik, Universit{\"a}t Regensburg,
             D-93040 Regensburg, Germany}
\author{Jens Siewert}
\affiliation{Institut f{\"u}r Theoretische Physik, Universit{\"a}t Regensburg,
             D-93040 Regensburg, Germany}
\affiliation{Dipartimento di Metodologie Fisiche e Chimiche per l'Ingegneria,
 Universita di Catania,  I-95125 Catania, Italy}
\author{Klaus Richter}
\affiliation{Institut f{\"u}r Theoretische Physik, Universit{\"a}t Regensburg,
             D-93040 Regensburg, Germany}

\date{\today}

\begin{abstract}
We present a linear-response theory for the thermopower
of a single-electron transistor consisting of a superconducting
island weakly coupled to two normal-conducting leads (NSN SET).
The thermopower shows oscillations with the same periodicity as
the conductance 
and is rather sensitive to the size
of the superconducting gap $\Delta$. In particular, the
previously studied sawtooth-like shape of the thermopower for
a normal-conducting
single-electron device is qualitatively changed even for small gap
energies.
\end{abstract}


\maketitle


%


The transport properties of small conducting grains in the Coulomb
blockade (CB) regime have extensively been studied during the past
years. This regime is characterized by a new energy scale, the
so-called charging energy $E_c$ of the grain (see below).
The most prominent phenomenon is the occurence of 
CB oscillations in the low-temperature conductance of a small 
grain weakly coupled to the leads~\cite{Averin91}. 
Recently, thermoelectric effects in single-electron devices 
such as the thermopower have attracted growing interest~\cite{Beenakker92,Staring93,Dzurak97,Moeller01,Andreev01,Boese01,Matveev02,Turek02,Scheibner04,Koch04}. 
%
%
The thermopower
is related to the current that arises due to
a finite temperature difference
between the two leads \cite{Abrikosov88}.
It yields additional information about the kinetics of the system
as it measures the average energy of the electrons
carrying the current through the system. Therefore some type
of electron-hole asymmetry in the system
%
is necessary in order to observe a non-vanishing thermopower. 
%
%

In analogy to the CB oscillations of the conductance the
thermopower of a small grain
shows oscillations of the same
periodicity but with  sawtooth-like 
shape~\cite{Beenakker92,Staring93}.
In contrast 
to the conductance this dependence on
the external gate voltage is very sensitive to
the conditions under which the thermoelectric
transport occurs. This sensitivity
has been demonstrated, e.g., for the transition
from the sequential tunneling regime to the
cotunneling regime~\cite{Dzurak97,Turek02}.
Recently, the thermopower of open quantum dots
with strong coupling to the leads was investigated
\cite{Moeller01, Andreev01,Matveev02}. 
Further, the influence
of Kondo correlations in ultra-small quantum dots
on the thermoelectric effects was studied in 
Refs.~\cite{Boese01, Scheibner04}
while the thermopower of a molecule with internal 
degrees of freedom and weakly coupled to the leads was 
discussed in Ref.~\cite{Koch04}.

\begin{figure}[b]
 \resizebox{.28\textwidth}{!}{\includegraphics{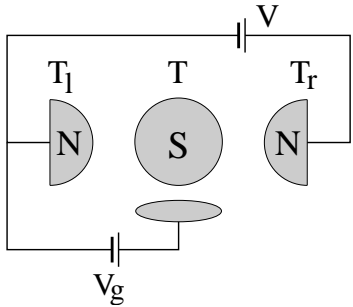}}
\caption{\label{fig:setup} 
  The NSN SET
  consists of a superconducting island (S) which
  is coupled to two
  normal-conducting leads (N) via tunnel barriers. The
  electrostatic potential of the island 
  can be controlled by the gate voltage $V_g$.
  The current through
  the system is due to the bias 
  voltage $V$ or a temperature
  difference $\Delta T = T_l-T_r$ between the two leads.
  To measure the thermopower $S = -V/\Delta T$
  as a function of the gate voltage $V_g$
  the bias $V$ is adjusted in such a way that the corresponding
  current exactly cancels the current 
  due to the temperature difference.
   }
\end{figure}

It is surprizing that, despite the enormous interest
in superconducting SETs,
the thermopower of such structures has not been
investigated yet.
In this work we study theoretically the thermopower
of an NSN SET, i.e., a small superconducting
island that is weakly coupled to normal-conducting leads
(cf.~Fig.~\ref{fig:setup}), in an experimentally
accessible regime.
We show that even for rather small superconducting
gaps (compared to the charging energy of the island)
the functional dependence of the thermopower on
the gate voltage is qualitatively changed while its
amplitude remains on the same order of magnitude.
This is in clear contrast to the corresponding
results for the conductance where the 
most pronounced effect is a suppression of the amplitude
with increasing gap size~\cite{Schon94}.

{\em Thermopower of single-electron devices --}
In the following we investigate the regime of single-electron
tunneling through a superconducting 
island with a charging energy $E_c$ that is large
compared to the temperature $T$, i.e. $E_c \equiv e^2/(2 C) \gg T$,
where $-e$ is the electron charge and $C$ the 
capacitance of the island. 
The temperature is assumed to be larger than
the crossover temperature for parity effects~\cite{Averin92}.
The electronic spectrum of the grain in the normal-conducting state
is assumed to be continuous and the conductances of the tunnel barriers
are much smaller than $e^2/h$. This implies that cotunneling processes
can be neglected and sequential tunneling dominates.
Taking into account the
external electrostatic potential $\phi \propto V_g$ 
imposed by the gate voltage $V_g$,
the total electrostatic energy of the island can be expressed
as
\begin{equation}
\label{eq:charging_energy}
E_n(\phi) = E_c \left( n^2 - 2 n \frac{C}{e}\phi \right) ,
\end{equation}
where $n$ is the number of excess electrons on the island.
To add one electron to the island
an energy $u_n(\phi) \equiv E_{n+1}(\phi) - E_n(\phi)$
is required. At low temperatures $T \ll E_c$ the
electronic transport is Coulomb-blocked. A current
flows only at potentials $\phi \approx \phi_n$ where 
$\phi_n$ is given by the condition 
$u_n(\phi_n) = 0$. With 
Eq.~(\ref{eq:charging_energy}) one finds
\begin{equation}
\label{eq:phi_n}
\phi_n = \frac{e}{C} \left( n + \frac{1}{2} \right).
\end{equation}

The current through the
device can be either due to
a transport voltage $V$ or a temperature difference $\Delta T=T_l-T_r$
between the two leads (see Fig.~1). Throughout this work we consider
the linear response regime, i.e., $e V / E_c \ll 1$ and
$\Delta T / T \ll 1$.

For zero temperature difference 
the linear response
to the voltage $V$ is given by the conductance $G_V$.
As a function of the potential $\phi$ it shows the
well-known CB peaks at $\phi = \phi_n$.
In the general case
with $\Delta T \neq 0$ the current 
is $I = G_V \, V + G_T \, \Delta T$.
The thermopower of the system is defined by the ratio of
voltage and temperature difference for 
vanishing current, i.e.,
\begin{equation}
\label{def:thermopower}
 S \equiv - \lim_{\Delta T \to 0} \left. \frac{V}{\Delta T} \right|_{I=0}
 = \frac{G_T}{G_V} .
\end{equation}

\begin{figure}
\centerline{
\epsfxsize=0.4\textwidth
\epsfbox{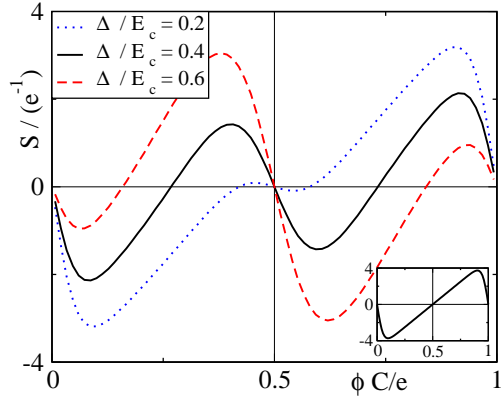}}
\caption{
\label{fig:s_dxxx_b10}
 Thermopower $S(\phi)$ of the NSN setup with $\Delta < E_c$ 
 for different values
 of the superconducting gap $\Delta$ at a temperature $T/E_c = 0.1$.
 Note that the conductance peak would be located at
 $\phi = \phi_0 \equiv 0.5 e/C$ [cf.~Eq.~(\protect\ref{eq:phi_n})]. 
 The inset shows the corresponding result for the NNN SET.}
\end{figure}

Following Matveev~\cite{Matveev00} 
a rather intuitive interpretation of the
CB oscillations of the thermopower can be given
in terms of the average energy $\langle \xi \rangle$
of the electrons that carry the current through the system
\begin{equation}
\label{eq:average_energy}
S = - \frac{\langle \xi \rangle}{e T} \ \ .
\end{equation}
Here, we briefly review the argument in the NNN case
(normal-conducting island) considering a two-state 
approximation (valid for \mbox{$T \ll E_c$}).
If in equilibrium there are $n$ electrons on the grain two different
transport cycles are possible: one can first add an electron and
then remove it again [$n \to (n+1) \to n$] or vice versa
[$n \to (n-1) \to n$]. In the first case the average energy
is given by the difference of charging energies,
$\langle \xi \rangle = (E_{n+1} - E_n)/2$, while in the
second case it is $\langle \xi \rangle = (E_{n} - E_{n-1})/2$.
The potential $\phi$ determines which of the processes
is more likely to occur. For example, at $\phi \gtrsim \phi_n$
the probability of having $n$ or $n+1$ electrons in the
grain is higher than the probability for $n-1$ electrons.
Thus, the first process dominates, and according to
Eq.~(\ref{eq:average_energy}) one finds the well-known
sawtooth behavior 
\begin{equation}
\label{eq:NNN_S}
 S_{\rm NNN} (\phi) = - \frac{u_n(\phi)}{2 e T}
 \quad {\rm for} \quad C |\phi - \phi_n| / e \, < \, \frac{1}{2} \  ,
\end{equation}
see inset Fig.~\ref{fig:s_dxxx_b10}.
The extrema of the sawtooth at \mbox{$e|\phi-\phi_n| \approx E_c$}
are rounded due to the finite temperature.


{\em Thermopower of  NSN SET --}
In the following we consider the case of a superconducting
grain with a gap $\Delta<E_c$. 
%
In a stationary state the currents through the left and the right
tunnel junction are equal, i.e., $I = I_l = I_r$.
In order to evaluate the thermopower we make use
of relation (\ref{def:thermopower}) and calculate 
the linear response of the current to a voltage $V$ or
a temperature difference $\Delta T = T_l-T_r$. According to the
``orthodox theory''~\cite{Averin91} the
current through the system can be written as
\begin{equation}
\label{eq:current}
I =  - e \sum_n P_n \left[ \Gamma_r^{\, n\to n-1} - \Gamma_r^{\, n\to n+1} 
                   \right]
\end{equation}
where $P_n$ is the stationary probability for finding $n$ electrons
on the island, $\Gamma_r^{n \to n-1}$ is the tunneling rate of an electron
from the island to the right lead, and $\Gamma_r^{n\to n+1}$ denotes the
tunneling rate from the right lead to the island.
The rates $\Gamma_{r}$ and correspondingly $\Gamma_{l}$ can be written
in terms of the Fermi-function $f(x)\equiv 1/[1+\exp (x)]$ as
\begin{widetext}
\begin{eqnarray}
\label{eq:rate_out}
 \Gamma_{r,l}^{\, n+1 \to n} =
 2 \frac{G_{r,l}}{e^2} & & \int\limits_\Delta^\infty
 d E \frac{E}{\sqrt{E^2-\Delta^2}} 
 \times \nonumber \\ 
 & & \left[
 f\left(-\frac{E \pm e V/2 + u_n}{T_{r,l}} \right) \,
 f\left(\frac{E}{T} \right) +
 f\left(\frac{E \mp e V/2 - u_n}{T_{r,l}} \right) \,
 f\left(-\frac{E}{T} \right)\right]  \\
\label{eq:rate_in}
 \Gamma_{r,l}^{\, n\to n+1} = 
 2 \frac{G_{r,l}}{e^2} & & \int\limits_\Delta^\infty
 d E  \frac{E}{\sqrt{E^2-\Delta^2}} 
 \times \nonumber \\
 & & \left[
 f\left(\frac{E \pm e V/2 + u_n}{T_{r,l}} \right) \,
 f\left(- \frac{E}{T} \right) +
 f\left(\frac{-E \pm e V/2 + u_n}{T_{r,l}} \right) \,
 f\left(\frac{E}{T}\right) \right] \ \ .
\end{eqnarray}
\end{widetext}
Here, $E$ is the energy of the quasiparticles in the superconductor,
and $G_{r,l}$ is the conductance of the right and left
tunnel junction, respectively (see, e.g., Ref.~\cite{Abrikosov88}). 
The first term in Eq.~(\ref{eq:rate_out})
corresponds to the annihilation of a quasiparticle while the second
term yields the contribution due to the creation of a quasiparticle.

The probabilities $P_n$ in Eq.~(\ref{eq:current}) can be obtained from
the stationary solution of a kinetic equation. They obey the
relation \cite{Beenakker92}
\begin{equation}
\label{eq:p_n}
P_{n+1} = \frac{\Gamma_l^{\, n \to n+1} + \Gamma_r^{\, n \to n+1}}
               {\Gamma_l^{\, n+1 \to n} + \Gamma_r^{\, n+1 \to n}} \; P_n 
\end{equation}
with $\sum_n P_n = 1$. The set of Eqs.~(\ref{eq:current})--(\ref{eq:p_n})
allows us to calculate the transport coefficients $G_V$ and
$G_T$. Together with relation (\ref{def:thermopower}) the
thermopower $S$  can be
obtained as a function of the potential $\phi$
for different parameters $\Delta$ and $T$. The results
of our calculation are shown in Fig.~\ref{fig:s_dxxx_b10}.
Compared to the sawtooth-like behavior of the thermopower
in the NNN-case (see inset Fig.~\ref{fig:s_dxxx_b10})
the shape is significantly changed for gaps
$\Delta$ smaller than the charging energy $E_c$.
For increasing $\Delta \to E_c$ the extrema of $S(\phi)$
move clearly away from $\phi_n \pm e / (2 C)$.

We emphasize that the qualitative change of the thermopower 
as a function of the potential $\phi$ due to a finite gap
is much more pronounced than the changes in the
conductance where one merely finds a broadening of the Coulomb-blockade
peaks together with an overall exponential  suppression of the
current. Note however that the order of magnitude of the thermopower
remains the same independently of the gap size.
This behavior can be understood by 
analyzing Eqs.~(\ref{eq:current})--(\ref{eq:p_n}) in the 
low-temperature regime
where the major contribution
to the electronic transport is due to only two charge states.

{\em Two-state approximation --}
For low temperatures \mbox{$T\ll E_c$}
there are at most two probabilities
that assume a finite value for a given $\phi$, 
e.g., $P_0$ and $P_1$ for $C |\phi - \phi_0| / e < 1/2$.
All remaining probabilities
are exponentially small in the parameter $E_c / T$.
To first order in the perturbations $V$ and $\Delta T$
the current (\ref{eq:current}) can be expressed in terms of the
unperturbed probabilities $P_{0,1}^{(0)}$ and
the exact rates $\Gamma_{l,r}$, Eqs.~(\ref{eq:rate_out}) and (\ref{eq:rate_in}),
as
\begin{eqnarray}
\label{eq:two_state_current}
 I  =  \frac{e}{G_l+G_r} & \left[ P_0^{(0)}
 \left( G_l \Gamma_r^{0 \to 1} - G_r \Gamma_l^{0 \to 1} \right) \right.
 + \\ \nonumber
 & \left. P_1^{(0)} 
 \left( G_r \Gamma_l^{1 \to 0} - G_l \Gamma_r^{1 \to 0} \right)
 \right] .
\end{eqnarray}
The zeroth-order probabilities can be obtained from 
Eq.~(\ref{eq:p_n}) and the condition $P_0^{(0)}+P_1^{(0)} = 1$
which yields $P_{0,1}^{(0)} \approx f[\mp u_0(\phi)/T]$
if exponentially small corrections are neglected.

First, we discuss the thermopower for potentials
$\phi$ close to $\phi_0$, i.e.,
$C |\phi - \phi_0| / e < \Delta / (2 E_c)$ where
$|u_0(\phi)| < \Delta$.
In this range of $\phi$, the changes in the thermopower due to the
superconducting properties of the island
are most prominent, see Fig.~\ref{fig:s_dxxx_b10}. 
For gap energies that are not too small 
$\Delta \lesssim E_c$ 
we can neglect contributions that
are exponentially small in $\Delta / T$. Thus we
find from 
Eqs.~(\ref{eq:rate_out}), (\ref{eq:rate_in}), (\ref{eq:two_state_current})
together with relation (\ref{def:thermopower}) the
asymptotic result
\begin{equation}
\label{eq:result1}
 S(\phi) \approx - \frac{u_0(\phi)}{e T} \left(
 1 - \frac{\tilde{\Delta}(T)}{u_0(\phi)}
 \, \tanh \left[ \frac{u_0(\phi)}{2 T} \right] \right) \ .
\end{equation}
Here,
$\tilde{\Delta}(T) \equiv -\Delta \; K'_1(\Delta / T) / K_1(\Delta / T)$
where $K'_1(x)$ is the derivative of the Bessel function
$K_1(x)$.
In contrast to the NNN case we find $S=0$ not only for
$u_0(\phi) = 0$ but also at 
$|u_0(\phi)| \approx \tilde{\Delta}$, see Fig.~\ref{fig:s_dxxx_b10}.
Between these two zeros
the thermopower reaches its extrema at
$|u_{\rm max}| \approx 2 T^{-1} {\rm arccosh}{\sqrt{\tilde{\Delta}/(2 T)}}$.
This novel behavior that the slope of $S(\phi)$ changes its sign
at $\phi = \phi_0$ 
occurs even for small gap values $\tilde{\Delta} \sim 2 T$.
In the low-temperature limit $\Delta / T \gg 1$
the temperature-dependent ``effective gap'' $\tilde{\Delta}(T)$ 
in Eq.~(\ref{eq:result1}) 
is simply replaced by the constant gap $\Delta$.
On the other hand, Eq.~(\ref{eq:result1}) 
also reproduces
the limit $\Delta \to 0$ correctly as it gives
$\tilde{\Delta} \to T$ leading to the
NNN result of Eq.~(\ref{eq:NNN_S}).

Next we consider the thermopower in the $\phi$ range
\mbox{$\Delta < |u_0(\phi)| < E_c$}, i.e.,
$\Delta / (2 E_c) < C|\phi-\phi_0|/e < 1/2$.
In this case Eq.~(\ref{eq:two_state_current}) 
and Eq.~(\ref{def:thermopower}) approximately yield
\begin{equation}
\label{eq:result2}
 S(\phi) \approx  - \frac{u_0(\phi)}{2e T} \left( 1 -
 \left[\frac{\Delta}{u_0(\phi)} \right]^2 
 \frac{{\rm arcosh} \left[ \frac{|u_0(\phi)|}{\Delta} \right] }
  {\sqrt{1-\left[\frac{\Delta}{u_0(\phi)} \right]^2}} \right) .
\end{equation}
From this result we find $S=0$ for $|u_0(\phi)| = \Delta$
which agrees with Eq.~(\ref{eq:result1}) in the
low-temperature limit where $\tilde{\Delta} \to \Delta$.
The second term in Eq.~(\ref{eq:result2})
depends only weakly on $\phi$ but
gives an overall shift by $\Delta / 2$,
see Fig.~\ref{fig:s_dxxx_b10}.
Therefore, the dependence of $S$ on the potential $\phi$
is almost linear so that the extrema
for $u_0(\phi) \to \pm E_c$ are approximately
given by $S_{\rm max} \approx \mp (E_c-\Delta)/(2 e T)$,
respectively.
Similarly to the NNN case there is a thermal smoothing close
to the edge of the $\phi$ interval where $E_c - |u_0(\phi)| \lesssim T$.
This is because for these values of $\phi$
charge states with $n=-1$ or $n=2$ electrons on the
island become important. For small gaps
$\Delta \to 0$ also Eq.~(\ref{eq:result2}) 
reproduces the NNN result of Eq.~(\ref{eq:NNN_S}).

\begin{figure}[t]
 \resizebox{.35\textwidth}{!}{\includegraphics{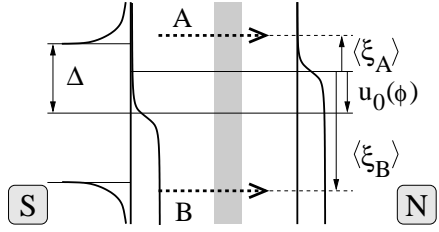}}
\caption{\label{fig:avg_energy} 
 Tunneling processes from the superconducting island to the
 right lead for $|u_0(\phi)|<\Delta$: 
 annihilation of a quasiparticle (A) and creation
 of a quasiparticle (B). 
 The corresponding average energies (measured from
 the Fermi energy of the lead) 
 are $\langle \xi_A \rangle > 0$ and $\langle \xi_B \rangle < 0$.
 Depending on which of the processes is dominant the sign
 of $S$ results according to Eq.~(\ref{eq:average_energy}).
   }
\end{figure}

{\em Interpretation of results in terms of average energy --}
The asymptotic results (\ref{eq:result1}) and (\ref{eq:result2}) can
be intuitively understood in terms of Eq.~(\ref{eq:average_energy})
by considering the average electron energy $\langle \xi \rangle$ 
of the dominating transport mechanism.
In Fig.~\ref{fig:avg_energy} we schematically present the
transport mechanisms for potentials
$\phi$ such that $|u_0(\phi)| < \Delta$. The two processes
corresponding to the rate $\Gamma_r^{1 \to 0}$, Eq.~(\ref{eq:rate_out}),
are shown. The Fermi level of the lead is chosen to be zero.
The average energy of the
electrons involved in process A is then 
$\langle \xi_A \rangle \approx u_0(\phi) + \Delta > 0$. 
On the other hand, process B involves the breaking of a Cooper pair:
one electron tunnels to the lead while the other remains
as a quasiparticle on the island.
Hence, the average energy of the outgoing electrons is
given by $\langle \xi_B \rangle \approx u_0(\phi) - \Delta < 0$.

As can be seen from Fig.~\ref{fig:avg_energy}, the two
processes are not equally likely to occur due to
the different occupation numbers in the island and the lead.
We can read off a low-temperature estimate for the probabilities
from Fig.~\ref{fig:avg_energy}: $p_A \propto \exp [-\Delta / T]$
and $p_B \propto \exp [-(\Delta - u_0)/T]$. Thus,
relation (\ref{eq:average_energy}) together with the
condition $p_A+p_B=1$ leads to the estimate
\begin{eqnarray}
\label{eq:estimate1_S}
 S &=& -\frac{1}{e T} \left( p_a \langle \xi_A \rangle
   + p_B \langle \xi_B \rangle \right) \nonumber \\
   &=& - \frac{1}{e T} \left( u_0(\phi) - 
   \Delta \tanh \left[ \frac{u_0(\phi)}{2 T} \right] \right) .
\end{eqnarray}
This corresponds precisely to the result~(\ref{eq:result1}) in the
low-temperature limit $\Delta / T \gg 1$ where
$\tilde{\Delta}(T) \to \Delta$.

If the potential $\phi$ increases further
such that \mbox{$u_0(\phi) < -\Delta$},
process A clearly dominates as $p_A \gg p_B$ and one can neglect
process B entirely. However, the tunneling of low-lying 
quasiparticle excitations is also strongly suppressed.
By including the 
%
%
energy dependence of the density of states in the superconductor
we find an estimate for the average energy 
$\langle \xi_A \rangle$ 
\begin{eqnarray}
 \langle \xi_A \rangle & \approx & u_0(\phi) + 
 \frac{\langle E_{(1)} \rangle}{\langle E_{(0)} \rangle} \quad {\rm with} \\
 \langle E_{(k)} \rangle & \equiv & \int\limits_\Delta^{-u_0(\phi)}
 dE \, E^k \, \frac{E}{\sqrt{E^2 - \Delta^2}} .
\end{eqnarray}
Including this estimate into relation~(\ref{eq:average_energy})
directly yields the result~(\ref{eq:result2}).

In conclusion, we have developed a theory for the thermopower
$S$ of a NSN SET with $\Delta < E_c$.
The numerically exact results are presented in Fig.~\ref{fig:s_dxxx_b10}
and the asymptotic low-temperature behavior is given in
Eqs.~(\ref{eq:result1}), (\ref{eq:result2}). We showed that
these results can be understood on the basis of the
average-energy interpretation Eq.~(\ref{eq:average_energy}).
We mention that the current-voltage characteristics of a
NSN SET in the relevant range of the parameters
discussed in this work have already 
been studied in experiments~\cite{Eiles93, Hergenrother94}.
It should be well within reach of present-day nanotechnology
to experimentally detect the sensitive dependence of the
thermopower on the gap size summarized in Fig.~\ref{fig:s_dxxx_b10}.

{\em Acknowledgments} 
The authors would like to thank R.~Fazio, R.~Scheibner, 
and Ch.~Strunk for helpful discussions.
Financial support from the DFG under contract Ri 681/5-1
and SFB 631 is gratefully acknowledged. JS is supported
by a Heisenberg fellowship of the DFG.


\end{document}